\documentclass[12pt]{article}

\newenvironment{wileykeywords}{\textsf{Keywords:}\hspace{\stretch{1}}}{\hspace{\stretch{1}}\rule{1ex}{1ex}}

\usepackage{amsmath,amssymb}
\usepackage{graphicx}
\usepackage{color}
\usepackage{dcolumn}
\usepackage{bm}
\usepackage[numbers,super,comma,sort&compress]{natbib}
\usepackage{multirow}
\usepackage{caption}
\usepackage{subcaption}

\usepackage{hyperref}

\definecolor{background-color}{gray}{0.98}

\title{Some challenges of diffused interfaces in implicit-solvent models}

\author{Mauricio Guerrero-Montero\thanks{Department of Mechanical Engineering, Universidad T\'ecnica Federico Santa Mar\'ia,Valpara\'iso, 2390123, Chile},
    Micha\l{} Bosy\thanks{School of Computer Science and Mathematics, Kingston University London, Penrhyn Road, Kingston upon Thames, KT1 2EE, UK} \thanks{ORCID: 0000-0003-2723-6913}, 
    Christopher D. Cooper\thanks{Department of Mechanical Engineering, Universidad T\'ecnica Federico Santa Mar\'ia,Valpara\'iso, 2390123, Chile}\thanks{Centro Cient\'ifico Tecnol\'ogico de Valpara\'iso, Universidad T\'ecnica Federico Santa Mar\'ia, Valpara\'iso 2390123, Chile} \thanks{ORCID: 0000-0003-0282-8998}}
\begin{document}

\maketitle

\begin{abstract}
The standard Poisson-Boltzmann model for molecular electrostatics assumes a sharp variation of the permittivity and salt concentration along the solute-solvent interface.
The discontinuous field parameters are not only difficult numerically, but also are not a realistic physical picture, as it forces the dielectric constant and ionic strength of bulk in the near-solute region.
An alternative to alleviate some of these issues is to represent the molecular surface as a diffuse interface, however, this also presents challenges.
In this work we analysed the impact of the shape of the interfacial variation of the field parameters in solvation and binding energy.
However we used a hyperbolic tangent function ($\tanh(k_p x)$) to couple the internal and external regions, our analysis is valid for other definitions.
Our methodology was based on a coupled finite element (FEM) and boundary element (BEM) scheme that allowed us to have a special treatment of the permittivity and ionic strength in a bounded FEM region near the interface, while maintaining BEM elsewhere. 
Our results suggest that the shape of the function (represented by $k_p$) has a large impact on solvation and binding energy.
We saw that high values of $k_p$ induce a high gradient on the interface, to the limit of recovering the sharp jump when $k_p\to\infty$, presenting a numerical challenge where careful meshing is key.
Using the FreeSolv database to compare with molecular dynamics, our calculations indicate that an optimal value of $k_p$ for solvation energies was around 3. 
However, more challenging binding free energy tests make this conclusion more difficult, as binding showed to be very sensitive to small variations of $k_p$. 
In that case, optimal values of $k_p$ ranged from 2 to 20.
\end{abstract}

\begin{wileykeywords}
\end{wileykeywords}

\clearpage



\makeatletter
  \renewcommand\@biblabel[1]{#1.}
  \makeatother

\bibliographystyle{apsrev}

\renewcommand{\baselinestretch}{1.5}
\normalsize

\clearpage

\section*{\sffamily \Large INTRODUCTION} 
Electrostatics plays a crucial role in molecular solvation, influencing both the structure and function of molecules. The solvent behaviour can be effectively described through electrostatic models which either explicitly account for the molecular structure, as seen in molecular dynamics (MD), or consider the solvent implicitly, as a continuum dielectric medium. Implicit solvent models, such as the Poisson-Boltzmann (PB) equation, are widely used to compute mean-field potentials, forces, and free energies of solvation and binding \cite{shen1995calculation,Baker2004,fu2022accurate}. 

The Poisson-Boltzmann equation results from applying Gauss' law of electrostatics to a continuum medium that contains free ions which screen the effect of charges. This is relevant to biological systems, where biomolecules are immersed in water with salt ions. Following Fig. \ref{fig:1_Iterfaz}, the solute is represented as a salt-free, low-dielectric cavity ($\Omega_m$) with point charges ($Q_i$) at the atoms' locations, inserted in a high-dielectric infinite ionic medium ($ie.$ the solvent, $\Omega_s$). 

\begin{figure}[h]
\centering
\includegraphics[scale=0.32]{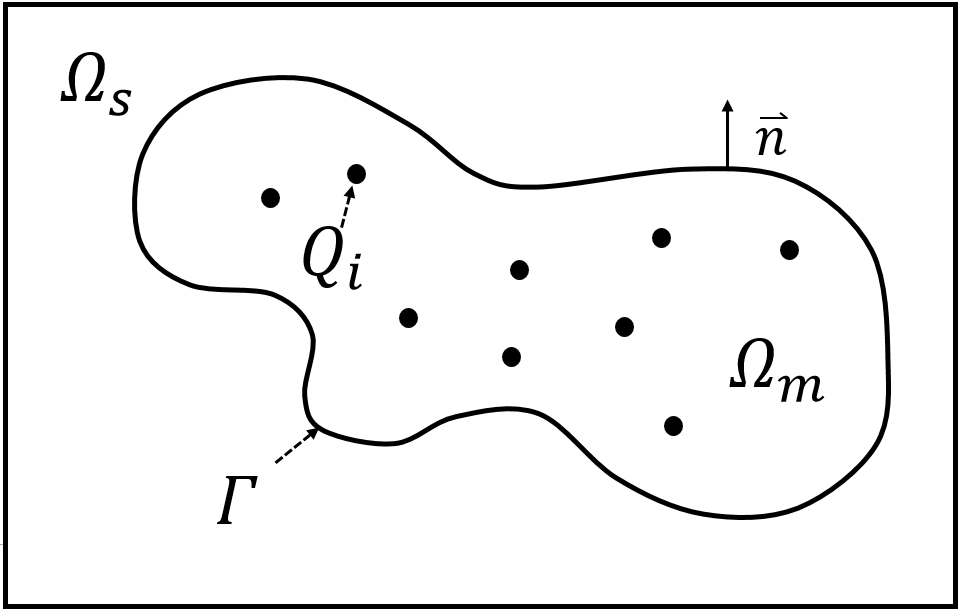}
\caption{Representation of a dissolved molecule ($\Omega_m$) in a Poisson-Boltzmann continuum solvent ($\Omega_s$).}
\label{fig:1_Iterfaz}
\end{figure}


The piece-wise constant distribution of the permittivity and salt concentration in Fig. \ref{fig:1_Iterfaz} may limit the applicability of the Poisson-Boltzmann equation. This sharp interface description presents two important drawbacks: 
\begin{enumerate}
    \item[(i)] using effective field variables from bulk is not a realistic physical picture \cite{bonthuis2011dielectric};
    \item[(ii)] the Dirac-delta description of the salt ions (equivalent to a radius equal to 0) overestimates their concentration near interfaces \cite{boschitsch2002fast,altman2009accurate}.
\end{enumerate}

Commonly used alternatives to the sharp jump are  diffuse interface approaches, which assume gradual and continuous transitions in the permittivity and ionic strength within a small region around the interface. For example, some couple the internal and external field parameters with linear \cite{xue2011three,bolcatto2001partially,sihvola1989polarizability,stern1978image}, cosine \cite{xue2011three,bolcatto2001partially,movilla2005image}, quasi-harmonic \cite{xue2017unified}, or hyperbolic tangent \cite{wang2021regularization,wang2022regularization} variations near the interface. These descriptions can be challenging to implement in realistic molecular geometries that are not spherical. 

Other models vary parameters throughout the whole solute region using Gaussian \cite{li2013dielectric, chakravorty2018reproducing, li2014modeling, wang2015pka} and Supergaussian \cite{panday2019reproducing, hazra2019super,wang2021regularization} functions, or a convolution of a Gaussian and the Heaviside function --- namely, the Gaussian Convolution Surface (GCS)~\cite{shao2023convergence}. 
These are usually preferred because they are easier to implement than the near-surface variations previously described. Nevertheless, they result in dielectric maps that are very similar: near-constant values in the internal region, and a smooth transition to the outside region.

The use of quasi-harmonic distributions has shown more accurate energetic predictions compared to its linear or cosine counterparts~\cite{xue2017unified}. On the other hand, the hyperbolic tangent function provides similar results to quasi-harmonic, ~\cite{wang2020regularization,wang2021regularization,wang2022regularization} descriptions for spheres. However, these models introduce new parameters for the space-varying descriptions that need to be determined. 

The Poisson-Boltzmann equation has been solved numerically using finite differences (FDM) \cite{GilsonETal1985,BakerETal2001,rocchia2001extending,boschitsch2011fast,JurrusETal2018}, finite elements (FEM) \cite{cortis1997numerical,chen2007finite,xie2007new,bond2010first}, boundary elements (BEM) \cite{shaw1985theory,yoon1990boundary,juffer1991electric,boschitsch2002fast,LuETal2006,geng2013treecode,cooper2014biomolecular,search2022towards}, (semi) analytical \cite{felberg2017pb,siryk2022arbitrary,jha2023linear,jha2023domain}, and hybrid approaches \cite{boschitsch2004hybrid,ying2018hybrid,bsbbbc2023coupling}.
BEM is a popular choice due to its accurate description of the point charges, molecular surface, and boundary conditions at infinity, however, it is limited to piece-wise constant variations of permittivity and salt concentration. A recent study \cite{bsbbbc2023coupling} addresses these limitations by combining BEM with FEM across the entire molecular region. This innovative approach enabled the application of a Gaussian distribution of permittivity. We plan to adopt a similar strategy by utilising FEM to simulate the diffuse interface, while employing BEM to represent the solvent and the remaining portion of the solute. 

The main goal of this work is to systematically study the impact of the diffuse interface's characteristics in the Poisson-Boltzmann predictions of solvation and binding free energy. We focus our attention on the near-surface changes in permittivity and salt concentration, through a hyperbolic tangent function. We choose this function because it requires fewer fitting parameters than, for example, quasi-harmonic functions, however, the analysis can be extended to other distributions, as discussed in the next section. 

\section*{\sffamily \Large METHODOLOGY}

\subsection*{\sffamily \large The linearised Poisson Boltzmann Equation with a sharp interface} 
\label{sec:sharp_inter}
In linearised form, the Poisson-Boltzmann equation (LPBE) reads
\begin{equation} 
    \label{eq:EcGeneral_LPBE}
    -\nabla\cdot(\epsilon(\mathbf{x})\nabla\phi(\mathbf{x}))+\bar{\kappa}^{2}(\mathbf{x})\phi(\mathbf{x})= \rho(\mathbf{x}),
\end{equation}
where $\epsilon$ corresponds to the relative permittivity, $\rho$ is the charge density, $\bar{\kappa}$ is the modified inverse of the Debye length, and $\phi$ is the electrostatic potential. Eq. \eqref{eq:EcGeneral_LPBE} is a good approximation of the electrostatic potential for low-to-moderate values of charge and salt concentration, like the ones found in proteins under physiological conditions.

Fig. \ref{fig:1_Iterfaz} shows the simplest application of the LPBE to molecular solvation, with the field parameters varying sharply through the molecular surface ($\Gamma$) as follows 
\begin{equation} 
\begin{matrix}
\epsilon(\mathbf{x})
\left\{\begin{matrix} 
     \epsilon_{m} & \mathbf{x}\in\Omega_{m} \\
     \epsilon_{s} & \mathbf{x}\in\Omega_{s}
\end{matrix}\right.
& \bar{\kappa}^{2}(\mathbf{x})
\left\{\begin{matrix} 
     0 & \mathbf{x} \ \in \ \Omega_{m} \\
     \epsilon_{s}\kappa_{s}^{2} & \mathbf{x}\in\Omega_{s} 
\end{matrix}\right. \\
\phi(\mathbf{x})
\left\{\begin{matrix} 
     \phi_{m}(\mathbf{x})  &  \mathbf{x}\in\Omega_{m} \\
     \phi_{s}(\mathbf{x})  &  \mathbf{x}\in\Omega_{s}  
\end{matrix}\right.
&\rho(\mathbf{x})
\left\{\begin{matrix}      \sum_{i=1}^{n_{c}}Q_{i}\delta_{\mathbf{x}_{i}}\left(\mathbf{x}\right) & \mathbf{x}\in\Omega_{m} \\
     0  &  \mathbf{x}\in\Omega_{s}  
\end{matrix}\right. 
\end{matrix}
\label{eq:Variables_1Superficie}
\end{equation}
Eq. \ref{eq:Variables_1Superficie} allows us to write the LPBE of Eq. \ref{eq:EcGeneral_LPBE} as a system of partial differential equations with interface conditions in $\Gamma$
\begin{equation} 
\label{eq:pbe_simple}
\left\{\begin{matrix} 
     -\epsilon_{m}\nabla^{2} \phi_{m}(\mathbf{x})=\sum_{i=1}^{n_{c}}Q_{i}\delta\left (\mathbf{x}-\mathbf{x}_{i} \right ) & \mathbf{x} \ \in \ \Omega_{m} \\[2mm]
     -\nabla^{2}\phi_{s}(\mathbf{x})+\kappa_{s}^{2}\phi_{s}(\mathbf{x})=0 & \mathbf{x} \ \in \ \Omega_{s} \\[2mm]
     \phi_{m}(\mathbf{x})= \phi_{s}(\mathbf{x}) & \mathbf{x} \ \in \ \Gamma \\[2mm]
     \epsilon_{m}\partial_n \phi_{m}(\mathbf{x})= \epsilon_{s}\partial_n \phi_{s}(\mathbf{x}) & \mathbf{x} \ \in \ \Gamma 
\end{matrix}\right. ,
\end{equation}
where expression $\partial_n := \tfrac{\partial}{\partial \mathbf{n}}$ refers to the  derivative normal to $\Gamma$.

\subsection*{\sffamily \large A modified description near the molecular surface}
\label{sec:diffuse_interface}

The sharp interface in Eq. \eqref{eq:Variables_1Superficie} assumes that the dielectric constant in the solute's vicinity is the same as in bulk (for water, that is usually 80), which is not a realistic physical picture. On one hand, molecules in the first few hydration shells interact with the solute atoms, limiting their freedom to move and reorient in response to an external electric field \cite{pal2004anomalous,sterpone2012magnitude,li2013dielectric}; on the other hand, water in confined spaces yields an unexpectedly low permittivity \cite{fumagalli2018anomalously}. This has served as inspiration to develop models that consider special treatments of the dielectric environment near the solute-solvent interface.

\begin{figure}[h]
\centering
\includegraphics[scale=0.31]{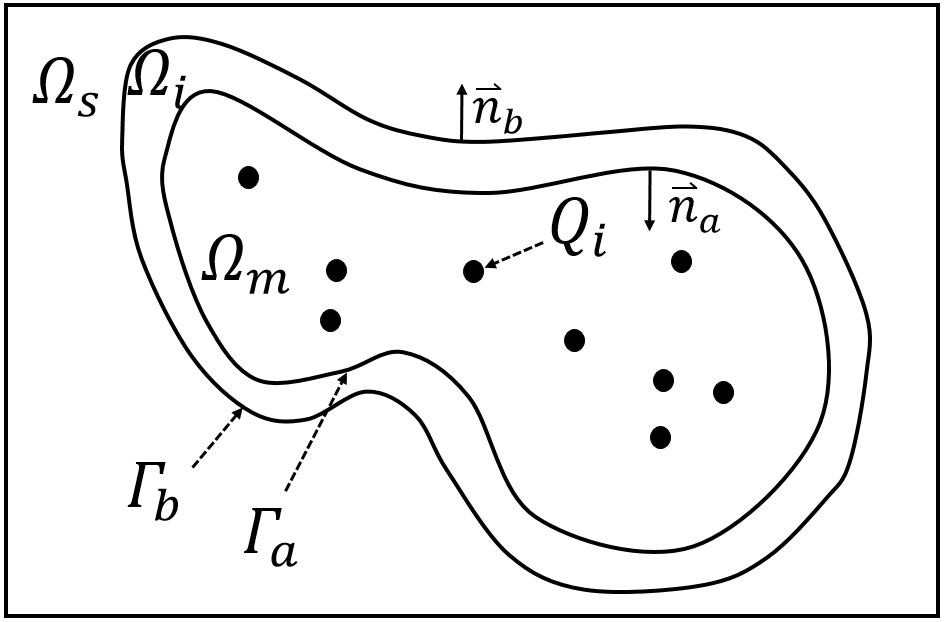}
\caption{Visualization of the domains for the two-surface model.}
\label{fig:2_Iterfaces}
\end{figure}

As an extension of the single-surface model in Fig \ref{fig:1_Iterfaz}, we can consider the two-surface model from Fig. \ref{fig:2_Iterfaces} that includes an interface layer region ($\Omega_i$) \cite{altman2009accurate}, where distinct values of $\epsilon$ and ionic concentration can be incorporated. This approach is similar to other well-known methodologies, such as considering ion-exclusion layers, aimed at displacing ions from the solute to alleviate the absence of steric effects in salt. We allow $\epsilon$ and $\kappa$ to vary in space in $\Omega_i$, as
\begin{equation} 
\label{eq:Variables_2Superficie}
\begin{matrix}
\epsilon(\mathbf{x})
\left\{\begin{matrix} 
     \epsilon_{m} & \mathbf{x}\in\Omega_{m} \\ 
     \epsilon_{i}(\mathbf{x}) & \mathbf{x}\in\Omega_{i} \\ 
     \epsilon_{s} & \mathbf{x}\in\Omega_{s}
\end{matrix}\right.
& \bar{\kappa}^{2}(\mathbf{x})
\left\{\begin{matrix} 
     0 & \mathbf{x} \ \in \ \Omega_{m}  \\  
     \bar{\kappa_{i}}^{2}(\mathbf{x}) & \mathbf{x}\in\Omega_{i}  \\  
     \epsilon_{s}\kappa_{s}^{2} & \mathbf{x}\in\Omega_{s} 
\end{matrix}\right. \\
\phi(\mathbf{x})
\left\{\begin{matrix} 
     \phi_{m}(\mathbf{x})  &  \mathbf{x}\in\Omega_{m} \\ 
     \phi_{i}(\mathbf{x})  &  \mathbf{x}\in\Omega_{i} \\ 
     \phi_{s}(\mathbf{x})  &  \mathbf{x}\in\Omega_{s}  
\end{matrix}\right.
& \rho(\mathbf{x})
\left\{\begin{matrix} 
     \sum_{i=1}^{n_{c}}Q_{i}\delta_{\mathbf{x}_{i}}\left(\mathbf{x}\right) & \mathbf{x}\in\Omega_{m} \\ 
     0  &  \mathbf{x}\in\Omega_{i} \\ 
     0  &  \mathbf{x}\in\Omega_{s}  
\end{matrix}\right.
\end{matrix}
\end{equation}
This results in the following system of partial differential equations
\begin{equation} 
\label{eq:pbe_vp}
\left\{\begin{matrix} 
     -\epsilon_{m}\nabla^{2} \phi_{m}(\mathbf{x})=\sum_{i=1}^{n_{c}}Q_{i}\delta\left (\mathbf{x}-\mathbf{x}_{i} \right ) & \mathbf{x} \ \in \ \Omega_{m} \\[2mm]
     -\nabla\cdot(\epsilon_{i}(\mathbf{x})\nabla\phi_{i}(\mathbf{x}))+\bar{\kappa_{i}}^{2}(\mathbf{x})\phi_{i}(\mathbf{x})
     =0 \ & \mathbf{x} \ \in \ \Omega_{i} \\[2mm] 
     -\nabla^{2}\phi_{s}(\mathbf{x})+\kappa_{s}^{2}\phi_{s}(\mathbf{x})=0 & \mathbf{x} \ \in \ \Omega_{s} \\[2mm]
     \phi_{m}(\mathbf{x})= \phi_{i}(\mathbf{x}) & \mathbf{x} \ \in \ \Gamma_{a} \\[2mm]
     \epsilon_{m}\partial_n^a \phi_{m}(\mathbf{x})= \epsilon_{i}(\mathbf{x})\partial_n^a \phi_{i}(\mathbf{x}) & \mathbf{x} \ \in \ \Gamma_{a} \\[2mm]
     \phi_{i}(\mathbf{x})= \phi_{s}(\mathbf{x}) & \mathbf{x} \ \in \ \Gamma_{b} \\[2mm]
     \epsilon_{i}(\mathbf{x})\partial_n^b \phi_{i}(\mathbf{x})= \epsilon_{s}\partial_n^b \phi_{s}(\mathbf{x}) & \mathbf{x} \ \in \ \Gamma_{b}
\end{matrix}\right. ,
\end{equation}
where $\Gamma_{a}$ and $\Gamma_{b}$ are the inner and outer surfaces of $\Omega_i$, with normal vectors according to Fig. \ref{fig:2_Iterfaces}. Following Fig. \ref{fig:Radio_de_Prueba}, $\Gamma_a$ corresponds to the solvent excluded surface (SES)~\cite{connolly1983analytical},
 and $\Gamma_b$ is a surface 3 \AA  (1 water diameter) away from the SES. The solvent accessible surface (SAS)~\cite{lee1971interpretation} is located in the middle between $\Gamma_a$ and $\Gamma_b$.

 \begin{figure}[h]
 \centering
 \includegraphics[scale=0.38]{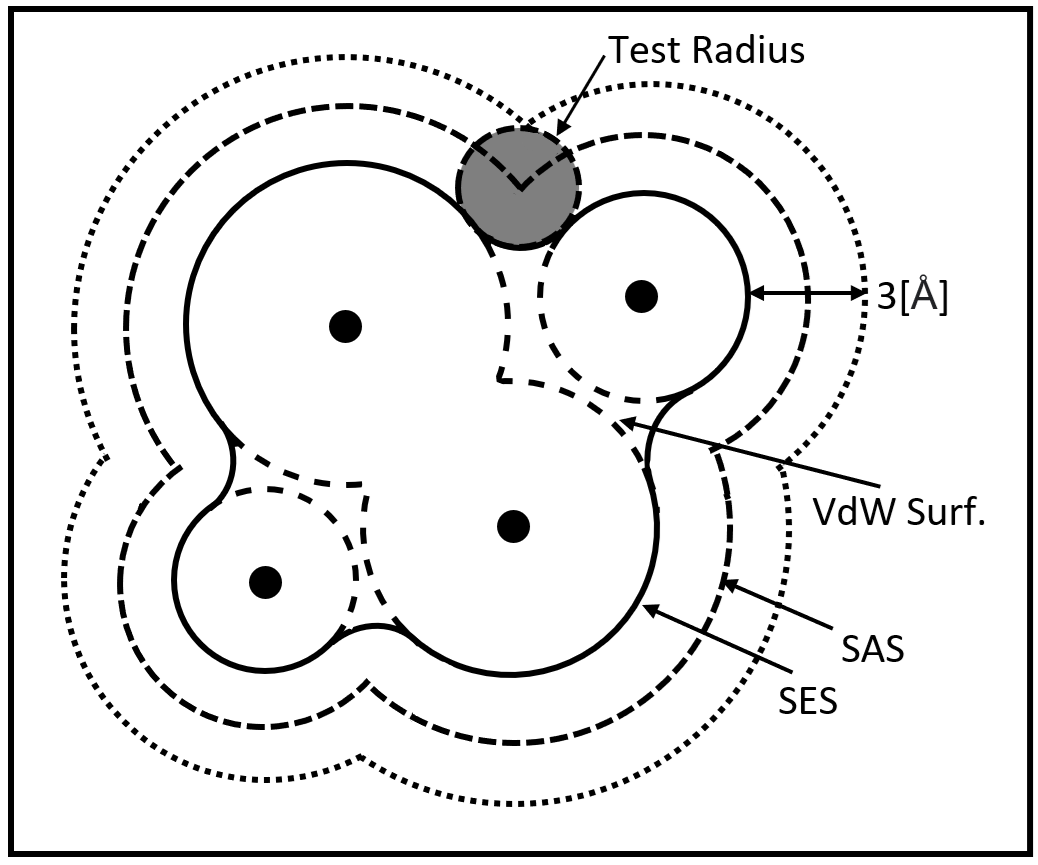}
 \caption{Visualization of the different surface definitions.}
 \label{fig:Radio_de_Prueba}
 \end{figure}

 We are considering the following functional form of the field parameters %
\begin{align*}
    \epsilon_i(\mathbf{x})=S(\mathbf{x},k_{p})\epsilon_m +(1-S(\mathbf{x} ,k_{p}))\epsilon_s \\
    \bar{\kappa_{i}}^{2}(\mathbf{x}) = (1-S(\mathbf{x} ,k_{p}))\epsilon_{s}\kappa_{s}^{2} 
\end{align*}
where the surface function $S$ can be smeared using a hyperbolic tangent ~\cite{wang2022regularization}
\begin{align}\label{eq:tanh}
    S(\mathbf{x},k_{p}) = \frac{1}{2}-\frac{1}{2}\tanh\left (k_{p}\left (\alpha(\mathbf{x})-\tfrac{1}{2} \right )\right ),
\end{align}
or the Gaussian Convolution Surface (GCS)~\cite{shao2023convergence} 
\begin{align*}
    S(\mathbf{x},\sigma) =\int_{\Omega_{i}}K(\mathbf{x}-\mathbf{t})d\mathbf{t} = \int_{\Omega_{i}}\frac{e^{-\frac{(\mathbf{x}-\mathbf{t})^{2}}{2 \sigma^{2}}}}{\sigma \sqrt{2 \pi}}d\mathbf{t}.
\end{align*} 
It is worth noting that the Gaussian approach yields an error function-type distribution
\begin{align*}
    S(\mathbf{x},\sigma) =\frac{1}{2}-\frac{1}{2}erf\left ( \frac{\alpha(\mathbf{x})-\tfrac{1}{2}}{\sigma\sqrt{2}} \right ).
\end{align*}
The definitions of both smeared surface functions are not only closely related but are also equivalent, as the hyperbolic tangent is an accurate approximation of the error function~\cite{mca27010014}. Consequently, while this paper focuses solely on the hyperbolic tangent formulation, the analysis presented herein is equally applicable to the GCS results. The only difference would be in the domain of parameter $\sigma$ as compared to $\kappa_p$, however, the resulting surface function would be the same.


\subsubsection*{\sffamily \large A hyperbolic tangent variation}\label{sec:tanh}

\begin{figure}[h!]
\centering
\includegraphics[scale=0.50]{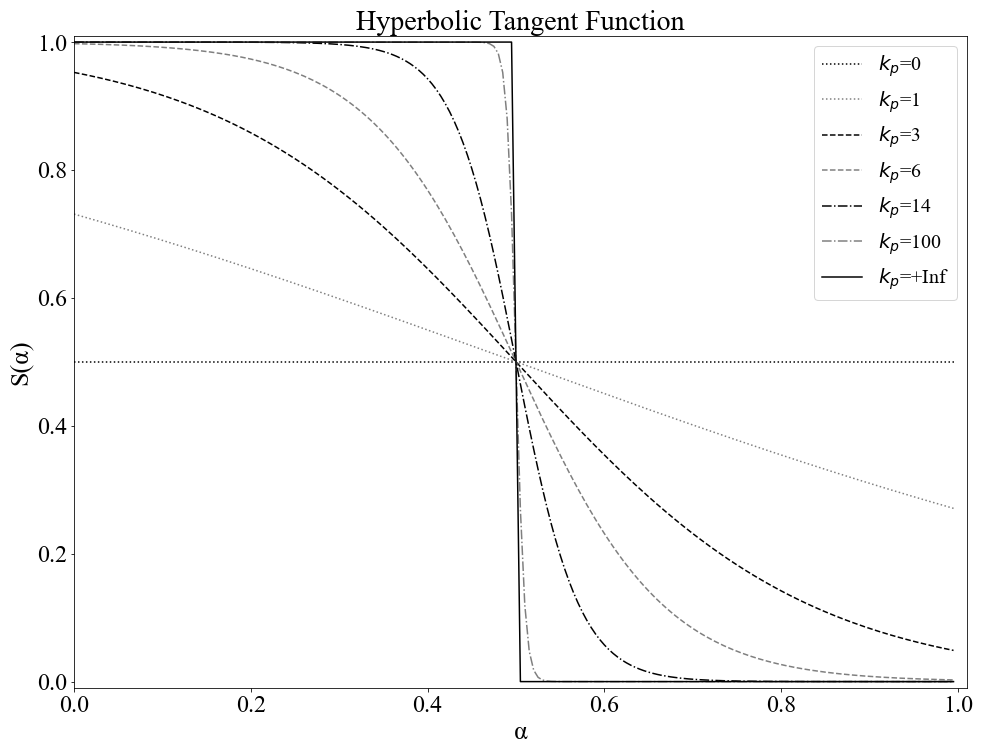}
\caption{behaviour of the hyperbolic tangent for different values of \(k_{p}\) and $\alpha$.}
\label{fig:TanH}
\end{figure}
\begin{figure}[h!]
\centering
\includegraphics[scale=0.35]{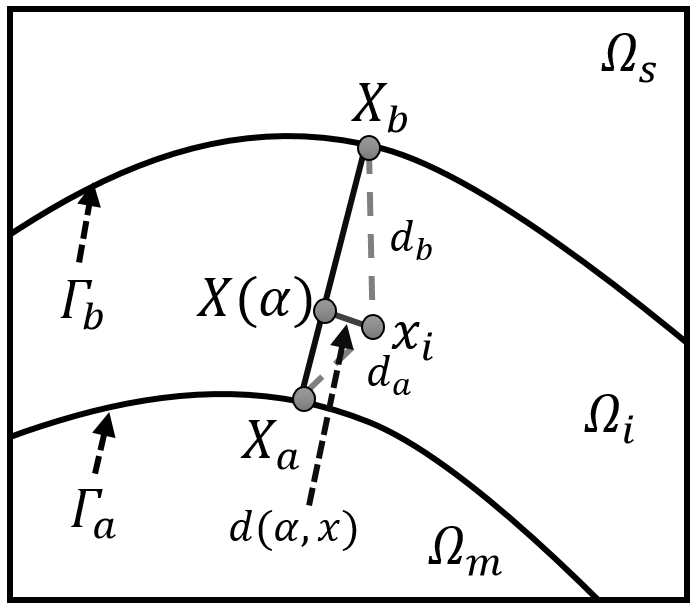}
\caption{Visualization of the scheme to calculate \(\alpha_{i}\) of a vertex \(i\) at position \(\mathbf{x}_{i}\) in the  intermediate domain.}
\label{fig:TanH_Calc_Mat}
\end{figure}

In Eq. \eqref{eq:tanh}, the dimensionless parameter \(k_{p}\)
corresponds to the slope at the function's transition (see Fig. \ref{fig:TanH}), and \(\alpha\) is a geometric parameter ranging from [0,1] that determines the position of the evaluation point ($x$) with respect to $\Gamma_a$ ($\alpha=0)$ and $\Gamma_b$ ($\alpha=1$), as shown by Fig. \ref{fig:TanH_Calc_Mat}.

Fig. \ref{fig:TanH} illustrates the behaviour of the hyperbolic tangent with respect to \(k_{p}\). The \(k_{p}=0\) case corresponds to a constant function across $\Omega_i$, with a value equal to the average of the solute and solvent. When \(k_{p}=+\infty\), the function recasts the sharp interface, this time, with a jump at \(\alpha(\mathbf{x})=0.5\), which corresponds to the SAS (see Fig. \ref{fig:Radio_de_Prueba}). 


Our approach requires \(\alpha\) for every vertex of the FEM mesh. Following Fig. \ref{fig:TanH_Calc_Mat}, for vertex $i$ located in $\mathbf{x}_i$ we first search for the nearest vertices from the meshes at $\Gamma_a$ and $\Gamma_b$, and define distances $d_a$ and $d_b$ as
\begin{align*} 
    d_a= min\left(\left \| \mathbf{X}_{a}-\mathbf{x}_{i} \right \|\right)  \quad  \quad d_b=min\left(\left \| \mathbf{X}_{b}-\mathbf{x}_{i} \right \|\right)
\end{align*}

Having found vertices \(\mathbf{X}_{a}\) and \(\mathbf{X}_{b}\), we create the following $\alpha$-dependent linear equation to find any point on the line connecting them:
\begin{equation} 
    X(\alpha)=\mathbf{X}_{a}+\alpha \left (\mathbf{X}_{b}-\mathbf{X}_{a}\right )
\label{eq:Ec.Lineal}
\end{equation}
%
%
The value $\alpha_i$ for $\mathbf{x}_i$ is that of the closest $X(\alpha)$ along the line defined by Eq. \eqref{eq:Ec.Lineal}. For $X(\alpha)$, the distance $d(\alpha,\mathbf{x})$ with $\mathbf{x}_i$ in Fig. \ref{fig:TanH_Calc_Mat} is minimal. The value of $d(\alpha)$ is defined as
\begin{equation} 
    d(\alpha_{i},\mathbf{x}_{i})=\left \| X(\alpha_{i} )-\mathbf{x}_{i} \right \|.
    \label{eq:Ec.Distancia}
\end{equation}
%
By simple optimization of Eq. \ref{eq:Ec.Distancia}, we can get the value of $\alpha_i$ as
\begin{align*} 
    \frac{\partial d(\alpha ,\mathbf{x})}{\partial \alpha }=0 \rightarrow \alpha_{i}(\mathbf{x}_{i})=-\frac{(\mathbf{X}_{a}-\mathbf{x}_{i})\cdot (\mathbf{X}_{b}-\mathbf{X}_{a})}{\left \| \mathbf{X}_{b}-\mathbf{X}_{a} \right \|^{2}}
\end{align*}

\begin{figure}[h]
\centering
\includegraphics[scale=0.44]{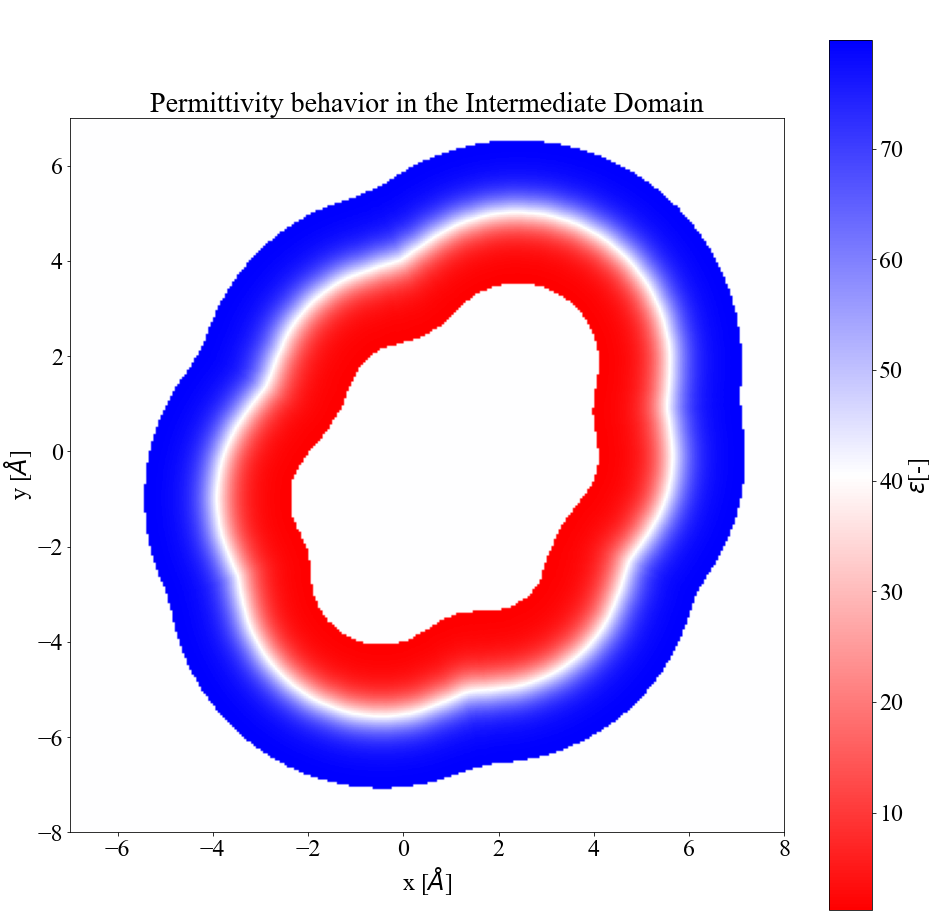}
\caption{Cross-section of the permittivity in the intermediate domain for cyclohexanol, using \(k_{p}=6\).}
\label{fig:Eps_Mobley}
\end{figure}

To avoid inconsistencies, we calculated $\alpha$ with three different surface mesh vertices $\mathbf{X}_a$ and $\mathbf{X}_b$, and used $\alpha_i$ as the average between them. Fig. \ref{fig:Eps_Mobley} shows an example of the resulting permittivity distribution on a cross-section of $\Omega_i$ for cyclohexanol, considering $k_p=6$ and a permittivity variation between $\epsilon_m$=1 and $\epsilon_s$=80. The SAS is clearly distinguishable as the white interface between the red and blue regions.

%
%

\subsection*{\sffamily \large FEM-BEM coupling}
\label{sec:FEM_BEM}

The FEM-BEM coupling is a powerful numerical technique used to solve various partial differential equations in science and engineering. The Johnson-Nédélec formulation, introduced in 1980~\cite{johnson1980coupling}, is the simplest formulation of FEM-BEM coupling. It was initially proposed for solving the linear Poisson-Boltzmann equation by Bosy {\it et al} \cite{bsbbbc2023coupling}. However, in our case, we are considering two interfaces, namely $\Gamma_a$ and $\Gamma_b$, instead of one. 

Since BEM cannot enforce the space-varying $\epsilon_i(\mathbf{x})$ and $\bar{\kappa}_i(\mathbf{x})$ in $\Omega_i$, we use FEM in that region. We begin by applying integration by parts to the second equation of~\eqref{eq:pbe_vp}. For every $v \in H_0^1(\Omega_i)$, we obtain:
\begin{align*}
\left(\epsilon_i \nabla \phi_i, \nabla v \right)_{\Omega_i} &+  \left(\kappa_i^2 \phi_i, \nabla v \right)_{\Omega_i} -  \left\langle  \epsilon_m\partial_n^a \phi_m , v \right\rangle_{\Gamma_a} -  \left\langle  \epsilon_s\partial_n^b \phi_s, v \right\rangle_{\Gamma_b} =  0, \nonumber 
\end{align*}
where $v$ is a test function and $\left\langle\varphi,v\right\rangle_\Gamma = \int_\Gamma \varphi(\mathbf{x})v(\mathbf{x})\,\mathrm{d}\mathbf{x}$ and $\left(\varphi,v\right)_{\Omega_i} = \int_{\Omega_i} \varphi(\mathbf{x})v(\mathbf{x})\,\mathrm{d}\mathbf{x}$ are the inner products on the surface and in the domain, respectively. 

For the external problem with BEM, we define the Dirichlet trace~\cite{MR2361676} 
\begin{align*}
\gamma:  H^1(\Omega_s) &\rightarrow H^{\frac{1}{2}}(\Gamma_b), & \gamma f(\mathbf{x}) & := \lim_{\Omega_s \ni \mathbf{y} \rightarrow \mathbf{x} \in \Gamma_b}  f(\mathbf{y}),
\end{align*}
and use the direct formulation from the third equation of Eq.~\eqref{eq:pbe_vp}  with $\lambda_s^b := \partial_n \phi_s$ to obtain
\begin{align*}
\tfrac{\phi_i}{2} - K_Y^b\gamma \phi_i + V_Y^b  \lambda_s^b & = 0.
\end{align*}
$V_Y$ and $K_Y$ are the single-layer and double-layer operators for the Yukawa (subscript $Y$) kernels. These are defined by
\begin{align*}
V_Y^b \varphi (\mathbf{x}) &= \oint_{\Gamma_b} g_Y(\mathbf{x},\mathbf{x}')\varphi(\mathbf{x}')d\mathbf{x}',\\
K_Y^b \varphi (\mathbf{x}) &= \oint_{\Gamma_b} \frac{\partial g_Y}{\partial\mathbf{n}'}(\mathbf{x},\mathbf{x}')\varphi(\mathbf{x}')d\mathbf{x}',
\end{align*}
where $\varphi(\mathbf{x})$ can be any distribution over $\Gamma_b$, and 
\begin{align*}
g_Y(\mathbf{x},\mathbf{x}')=\frac{e^{-\kappa|\mathbf{x}-\mathbf{x}'|}}{4\pi|\mathbf{x}-\mathbf{x}'|}
\end{align*}
is the corresponding free-space Green's function.

Similarly, the internal problem with BEM, we define the Dirichlet trace
\begin{align*}
\gamma:  H^1(\Omega_m) &\rightarrow H^{\frac{1}{2}}(\Gamma_a), & \gamma f(\mathbf{x}) & := \lim_{\Omega_m \ni \mathbf{y} \rightarrow \mathbf{x} \in \Gamma_a}  f(\mathbf{y}),
\end{align*}
and we use the direct formulation from the first equation of Eq.~\eqref{eq:pbe_vp} with $\lambda_m^a := \partial_n^a \phi_m$ to obtain
\begin{align*}
\tfrac{\phi_i}{2} - K_L^a\gamma \phi_i + V_L^a  \lambda_m^a & =  \frac1{4\pi\epsilon_m}\sum_{k=0}^{N_q}  \frac{Q_i}{|\mathbf{x}_{a} - \mathbf{x}_k|}.
\end{align*}
$V_L$ and $K_L$ are the single-layer and double-layer operators for the Laplace (subscript $L$) kernels. These are defined by
\begin{align*}
V_L^a \varphi (\mathbf{x}) &= \oint_{\Gamma_a} g_L(\mathbf{x},\mathbf{x}')\varphi(\mathbf{x}')d\mathbf{x}',\nonumber\\
K_L^a \varphi (\mathbf{x}) &= \oint_{\Gamma_a} \frac{\partial g_L}{\partial\mathbf{n}'}(\mathbf{x},\mathbf{x}')\varphi(\mathbf{x}')d\mathbf{x}',
\end{align*}
where $\varphi(\mathbf{x})$ can be any distribution over $\Gamma_a$, and 
\begin{align*}
g_L(\mathbf{x},\mathbf{x}')=\frac{1}{4\pi|\mathbf{x}-\mathbf{x}'|} 
\end{align*}
is the corresponding free-space Green's function.

Then, the coupling problem can be written as: \textit{Find $\lambda_m^a \in H^{-\frac{1}{2}}(\Gamma_a)$, $ \phi_i \in H^1(\Omega_i)$ and $\lambda_s^b \in H^{-\frac{1}{2}}(\Gamma_b)$ such that for all $\delta \in H^{-\frac{1}{2}}(\Gamma_a)$,  $v \in H^1(\Omega_1)$ and $\zeta \in H^{-\frac{1}{2}}(\Gamma_b)$},
\begin{align}
\label{eq:standard_fem_bem} \nonumber 
 \left\langle \left(\tfrac{1}{2} I - K_{L}^{a}\right) \gamma \phi_i, \delta \right\rangle_{\Gamma_a} + \left\langle V_{L}^{a}\lambda_m^a, \delta \right\rangle_{\Gamma_a}&= \left\langle \phi_c^a, \delta \right\rangle_{\Gamma_a}, \\[2mm] \nonumber
 \left(\epsilon_i \nabla \phi_i, \nabla v \right)_{\Omega_i} +  \left(\kappa_i^2 \phi_i, \nabla v \right)_{\Omega_i} \nonumber
 -  \left\langle  \epsilon_m\lambda_m^a , v \right\rangle_{\Gamma_a} -  \left\langle  \epsilon_s\lambda_s^b, v \right\rangle_{\Gamma_b} &=  0, \\[2mm] 
  \left\langle \left(\tfrac{1}{2} I - K_{Y}^{b}\right) \gamma \phi_i, \zeta \right\rangle_{\Gamma_b} + \left\langle V_{Y}^{b}\lambda_s^b, \zeta \right\rangle_{\Gamma_b} &=0.
\end{align}
where 
\begin{equation*}
   \phi_c^a := \frac1{4\pi\epsilon_m}\sum_{k=0}^{N_q}  \frac{Q_i}{|\mathbf{x}_{a} - \mathbf{x}_k|}.
\end{equation*}
The discrete problem in \eqref{eq:standard_fem_bem} can be written in the following blocked matrix form:
\begin{align}\label{eq:fembem_matrix}
    \begin{bmatrix}
    V_{L}^{a} & \frac{1}{2}\widetilde{M_{a}}-K_{L}^{a} & 0 \\[1mm]
    -\epsilon_{m}\widetilde{M_{a}}^T & \epsilon_{i}A -\kappa_{i}^{2}M  & -\epsilon_{s}\widetilde{M_{b}}^T \\[1mm]
    0 & \frac{1}{2}\widetilde{M_{b}}-K_{Y}^{b} & V_{Y}^{b}
    \end{bmatrix} \begin{bmatrix}
   \lambda_{m}^a \\[1mm] \phi_{i} \\[1mm] \lambda_{s}^b 
    \end{bmatrix} =\begin{bmatrix}
    \phi_{c}^a\\[1mm] 0 \\[1mm] 0
    \end{bmatrix}.
\end{align}
where 
\begin{equation*}
   \widetilde{M_{a}}^T = \left\langle  \lambda_m^a , v \right\rangle_{\Gamma_a}, \ \widetilde{M_{b}}^T = \left\langle  \lambda_s^b, v \right\rangle_{\Gamma_b},  \ M = \left(\phi_i, \nabla v \right)_{\Omega_i}.
\end{equation*}

\subsection*{\sffamily \large Energy calculations}
This section presents free energies of solvation and binding, which are common quantities of interest in PB solvers. Here we used them as the main evaluation metrics to asses our results. 

\subsubsection*{\sffamily \large Solvation energy}

The solvation energy corresponds to the free energy requred to dissolve a solute in a polarized solvent. In implicit solvents, this energy is typically divided into polar (electrostatic) and nonpolar components. The electrostatic energy is generally defined as:
\begin{equation} 
    \label{eq:Energia}
    E=\int_{\Omega }\rho(\mathbf{x})\phi(\mathbf{x})d\Omega 
\end{equation}  
%
where $\rho$ is the charge density. If we use the charge density in the solute from Eq. \ref{eq:Variables_2Superficie} and define the reaction potential ($\phi_r$) as the difference in electrostatic potential between the dissolved ($\phi_m$) and vacuum ($\phi_{coul}$) states ($\phi_{r} = \phi_m - \phi_{coul}$), the solvation energy is finally expressed as:
\begin{equation} 
    \label{eq:Esolv}
    \Delta G_{Sol}=\frac{1}{2}\sum_{i=1}^{n_{c}}Q_{i}\phi_{m}(\mathbf{x}_i). 
\end{equation}   
%


\subsubsection*{\sffamily \large Binding free energy}

\begin{figure}
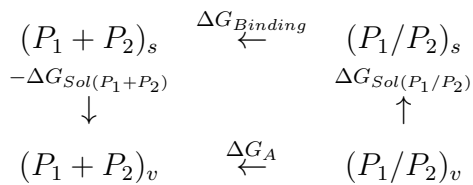
 
\centering
\begin{equation*}
    \begin{matrix}
    (P_{1}+P_{2})_{s} & \overset{\Delta G_{Binding} }{\leftarrow}  & (P_{1}/P_{2})_{s}\\ 
    \overset{-\Delta G_{Sol(P_{1}+P_{2})}}{\downarrow}  &  & \overset{\Delta G_{Sol(P_{1}/P_{2})}}{\uparrow} \\ 
    (P_{1}+P_{2})_{v} & \overset{\Delta G_{A} }{\leftarrow}  & (P_{1}/P_{2})_{v}
    \end{matrix}
\end{equation*}
    \caption{Thermodynamic cycle for binding energy. Top and bottom states are solvated (subscript $s$) and vacuum (subscript $v$), respectively. Left and right states are bound and unbound, respectively.}
    \label{fig:CicloTermodinamico}
\end{figure}

The binding free energy for a complex formed by the union of any two molecules corresponds to the minimum energy required to keep the complex structure together. We can calculate it using the thermodynamic cycle in Fig. \ref{fig:CicloTermodinamico} for molecules $P_1$ and $P_2$, and complex $P_1+P_2$. Considering that the Gibbs free energy depends only on the thermodynamic state (adding to zero in a closed cycle), the binding free energy ($\Delta G_{Binding}$ in Fig. \ref{fig:CicloTermodinamico}) is the difference of solvation free energies in bound and unbound states ($\Delta G_{P_1+P_2}$ and $\Delta G_{P_1/P_2}$ in Fig. \ref{fig:CicloTermodinamico}, respectively), plus the binding energy in vacuum ($\Delta G_A$ in Fig. \ref{fig:CicloTermodinamico}). $\Delta G_A$ corresponds to the difference in Coulomb energy between the isolated and bound structures:
\begin{align}   
    \Delta G_{A}&=E_{C,(P_{1}P_{2})_{v}}-E_{C, (P_{1}+P_{2})_{v}}\nonumber \\
    E_{C} &= \frac{1}{2}\frac{1}{\epsilon_{m} }\sum_{i=1}^{N}Q_{i}\sum_{j=1,i\neq j }^{N}\frac{Q_{j}}{R_{ij}}
\end{align}
Then, the electrostatic component of the binding free energy becomes:
\begin{align} \label{eq:bind_energy}  
    \Delta G_{Binding} = \Delta G_{Sol(P_{1}P_{2})}+E_{C,(P_{1}P_{2})_{v}} 
    -E_{C, (P_{1})_{v}}-E_{C, (P_{2})_{v}}
    -\Delta G_{Sol(P_{1})}-\Delta G_{Sol(P_{2})}
\end{align}

\section*{\sffamily \Large RESULTS AND DISCUSSION}
In this section, we analyse the characteristics of a diffuse interface using the hyperbolic tangent function, denoted as $\tanh$. Our analysis is divided into three parts. Firstly, we use a spherical domain to study the effect of different meshes in order to determine the most suitable one for the rest of our analysis. Next, we  calculate the solvation energy of nearly 500 molecules of the FreeSolv dataset \cite{mobley2014freesolv} to parameterize the $\tanh$ function. Finally, we study the impact of this parameterization on more challenging binding energy calculations, for 5 protein-protein and 2 protein-ligand complexes.

The results reported in this section were obtained from runs performed on a workstation with two Intel(R) Xeon(R) CPU E5-2680 v3 @ 2.50GHz with 12 cores each and 96 GB RAM memory.
The software was built using Bempp-cl 0.3.1 \cite{betcke2021bempp} and Dolfin from FEniCS 2019.1.0 \cite{logg2010dolfin}, running on Python 3.8.
We also used Trimesh \cite{trimesh}, MSMS \cite{sanner1996reduced}, Nanoshaper \cite{decherchi2013general}, PyGAMER \cite{lee2020open}, and PBJ \cite{search2022towards} for mesh generation and manipulation.  
All codes required to reproduce the results presented in this section are available in \url{https://github.com/bem4solvation/BEM-FEM-BEM_Coupling_with_Tangent_Hiperbolic}.

\subsection*{\sffamily \large Sphere}\label{}

As an initial test, we computed the solvation energy of a sphere of radius 5 \AA~ and a 3 \AA~ layer, containing three point charges of $q$=1$e$, positioned at (0,0,0) and (0,$\pm$0.5,0) \AA. The solute permittivity was $\epsilon_{m}=4$, while the solvent was set to $\epsilon_{s}=80$ with an inverse of length of Debye length $\kappa$=0.125 \AA$^{-1}$, corresponding to 150 mM of sodium chloride dissolved in water.

\begin{figure}
    \begin{subfigure}[b]{\linewidth}
    \centering
    \includegraphics[scale=0.36]{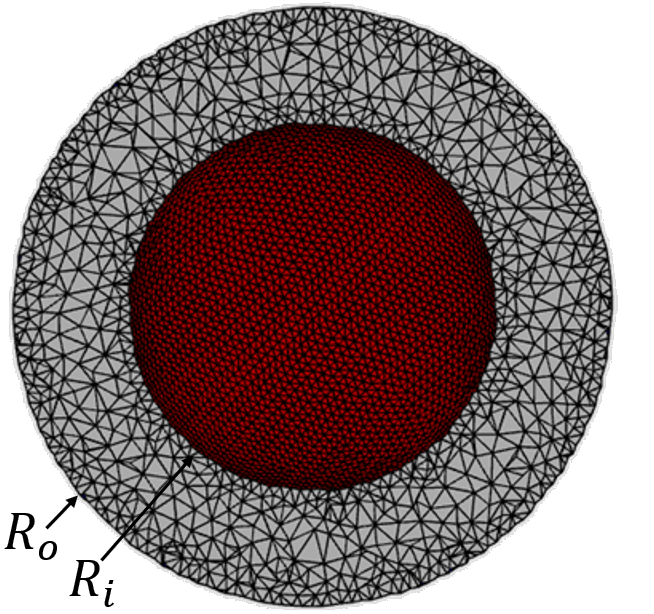}
    \caption{Restricted only by a growth factor of 1.1.}
    \label{fig:Esfera_SMI} 
    \end{subfigure} 
    \begin{subfigure}[b]{\linewidth}
    \centering
    \includegraphics[scale=0.36]{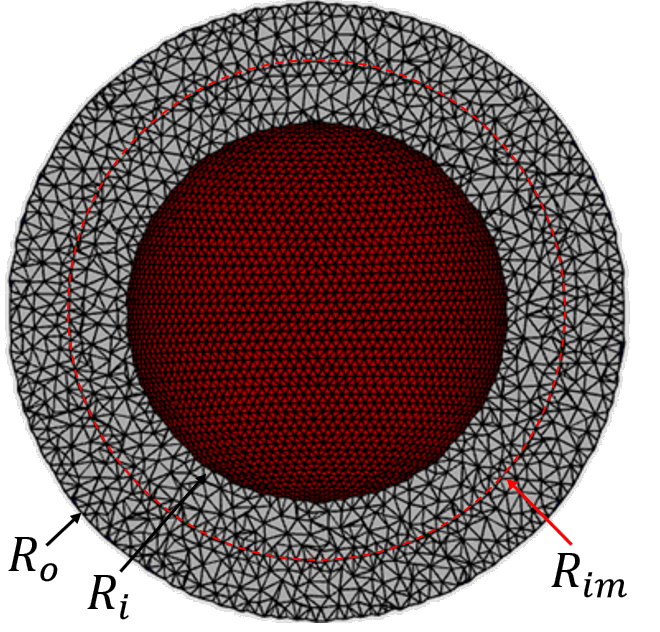}
    \caption{Using a intermediate surface mesh.}
    \label{fig:Esfera_CMI}
    \end{subfigure}
    \begin{subfigure}[b]{\linewidth}
    \centering
    \includegraphics[scale=0.41]{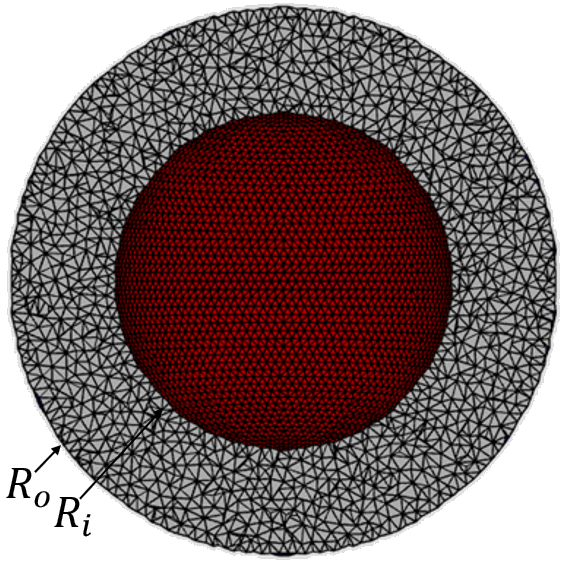}
    \caption{Using a element volume limit of 0.013 \AA$^3$.}
    \label{fig:Esfera_REG}
    \end{subfigure}
\caption{Cross sectional plane of the three volumetric meshes for a the spherical shell}
\end{figure}

The inner and outer surface meshes were generated with MSMS \cite{sanner1996reduced} using a density of 15 vertices per \AA$^2$, resulting in meshes with 4591 and 11924 vertices (9178 and 23844 elements), respectively. 
Starting from those surface meshes, we generated three FEM meshes using TetMesh through PyGAMER \cite{lee2020open}, limiting the growth of adjacent elements to less than 10\% (growth factor set to 1.1), for the following cases:
\begin{enumerate}
    \item Without any additional restrictions to the grid generation, resulting in a volume mesh with 36579 vertices. This mesh is shown in Fig. \ref{fig:Esfera_SMI}.
    \item Adding a third surface mesh between the internal and external ones of radius 6.5 \AA, also with 15 vertices per \AA$^2$. The resulting FEM mesh contained 50529 vertices, and is shown in Fig. \ref{fig:Esfera_CMI}.
    \item Limiting the volume of mesh elements to a maximum of 0.013 \AA$^3$, such that the number of elements is similar to the case with intermediate mesh. In this case, the FEM mesh had 50390 vertices, and is shown in Fig. \ref{fig:Esfera_REG}. Note that the number of elements of this mesh is very close to the case with the intermediate surface mesh.
\end{enumerate}




\begin{figure} [h]
\centering
\includegraphics[scale=0.50]{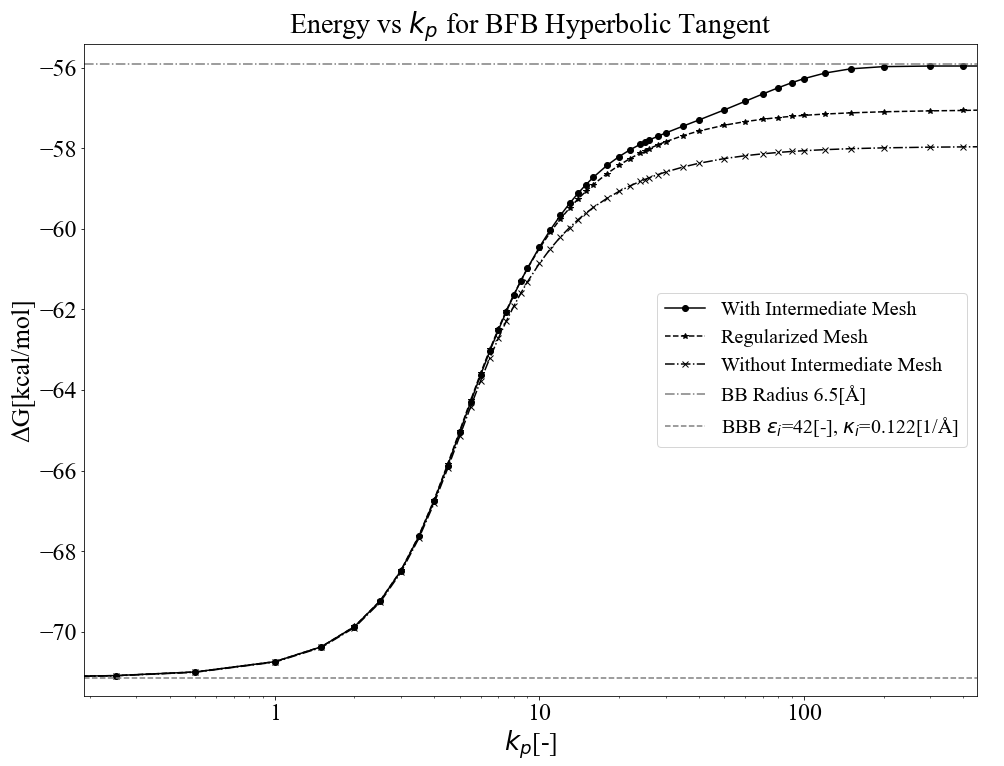}
\caption{Solvation energy of the sphere for different values of \(k_{p}\) in tanh of the BEM-FEM-BEM, using the three meshes detailed in Fig. \ref{fig:Esfera_REG}. BB and BBB refer to the reference calculations that couple BEM-BEM and BEM-BEM-BEM, respectively.}
\label{fig:Esfera_EvsKp}
\end{figure}

Fig. \ref{fig:Esfera_EvsKp} shows the solvation energy for different values of $k_p$ using the meshes detailed in Figs. \ref{fig:Esfera_SMI}, \ref{fig:Esfera_CMI}, and \ref{fig:Esfera_REG}. As a reference, Fig. \ref{fig:Esfera_EvsKp} includes BEM-BEM coupling results to which the $k_p=0$ and $k_p=\infty$ should converge. When $k_p=0$, the intermediate region has constant the parameters $\epsilon_i$=42 and $\kappa_i$=0.122 \AA$^{-1}$, solvable with BEM-BEM-BEM coupling, resulting in a solvation energy of -71.152 kcal/mol. If $k_p\to\infty$, there is a sharp interface where the radius is 6.5 \AA, and a single-surface BEM at the SAS is enough, yielding $\Delta G_{Solv}=-55.902$ kcal/mol. 

As expected from the behaviour of the hyperbolic tangent in Fig. \ref{fig:TanH}, having an appropriate mesh near the SAS, where the $\tanh$ function has its highest gradients, is key for accurate simulations. This is especially true for high values of $k_p$, where we see differences up to 2 kcal/mol between the meshes in Fig. \ref{fig:Esfera_CMI} and \ref{fig:Esfera_REG}, that have the same average density. Even though these differences may be negligible for solvation energy calculations, they can become important when evaluating differences of solvation energy, for example, binding energy (see Eq. \eqref{eq:bind_energy}). These results also highlight the large effect of the choice of $k_p$ in solvation energy, varying nearly 15 kcal/mol throughout Fig. \ref{fig:Esfera_EvsKp}.



\subsection*{\sffamily \large The FreeSolv Database}


To study the impact of the $\tanh$ variation in more realistic geometries, we computed the solvation energy of 494 molecules from the FreeSolv\footnote{\url{https://github.com/MobleyLab/FreeSolv}} database \cite{mobley2014freesolv}. Freesolv contains a set of small neutral organic molecules, with their corresponding solvation energies computed using molecular dynamics (MD), and divided in their polar and non polar components. The original PRMTOP, CRD, and MOL2 files from the database were transformed to PQR format using ParmEd\footnote{\url{https://parmed.github.io/ParmEd/html/index.html}}.


The creation of the inner, intermediate, and outer surface meshes were done through MSMS \cite{sanner1996reduced} forcing an average density of 8 vertices per Å$^{2}$. For the intermediate and outer surfaces we increased the van der Waals radii of each atom by 1.5 \AA  and 3 \AA, respectively. As for the volumetric mesh with TetMesh, it was generated with a growth factor of 1.1 on adjacent tetrahedra.


Regarding physical conditions, we set the the solute's relative permittivity to \(\epsilon_{m}=1\), while the solvent followed the condition of salt-free water, with permittivity values of \(\epsilon_{s}=80\) and an inverse length of \(\kappa=0\) \AA$^{-1}$.

\begin{figure}[h]
\centering
\includegraphics[scale=0.40]{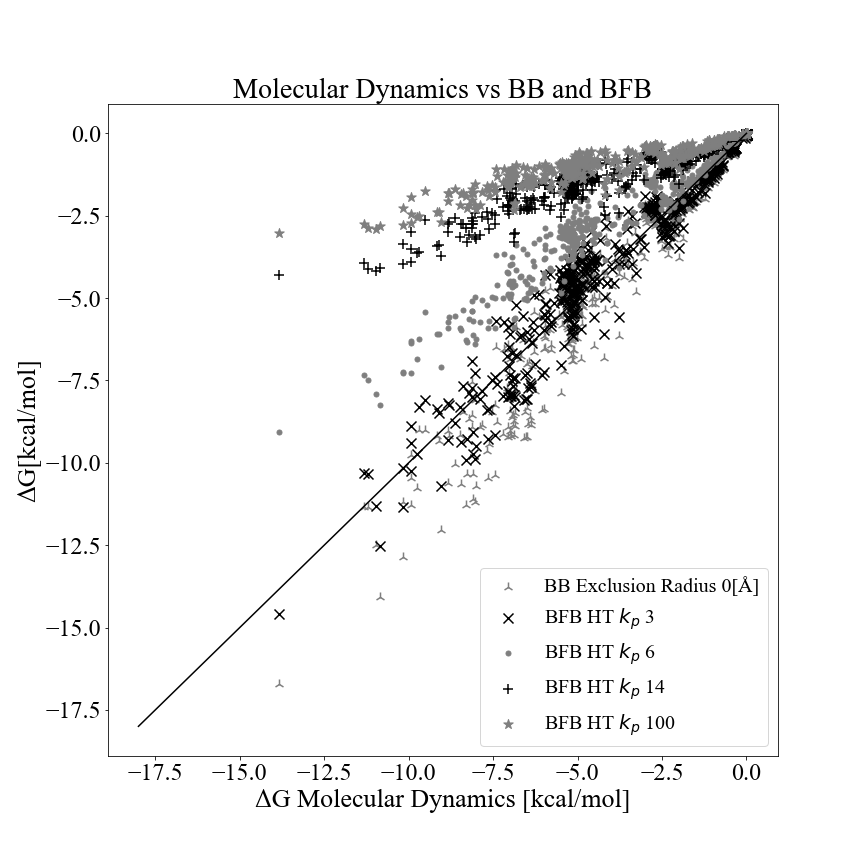}
\caption{Solvation energy of 469 molecules from the FreeSolv Database for different values of $k_p$ in tanh using BEM-FEM-BEM (BFB) and BEM-BEM (BB),  compared with MD.}
\label{fig:Mobley}
\end{figure}

The scatter plot in Fig. \ref{fig:Mobley} shows solvation energies computed with various values of \(k_{p}\) in the tanh for BEM-FEM-BEM, and the the standard BEM-BEM case on the SES (without VdW radius increment). These results are compared with those obtained from MD, where the 25 worst-performing molecules out of the 494 are excluded.
We can see that using a \(k_{p}\) value of 3 yields the closest results with respect to MD. This is confirmed in Table \ref{table:ECM}, where for different \(k_{p}\) values, the lowest mean squared error and the highest correlation coefficient are precisely found at $k_p$= 3 for this group of molecules. 

Fig. \ref{fig:Mobley} also shows how high values of $k_p$ underestimates the solubility, resulting in less favourable solvation free energies, which do not correlate with MD simulations. 
This is very sensitive at low values of $k_p$, but less so for higher values. 
For example, modifying $k_p$ from 3 to 6 has a large impact in solvation free energy, however, results with $k_p$=14 and $k_p$=100 are closer together. 
This sensitivity is also evident in Table \ref{table:ECM}.

\begin{table}
\centering
\footnotesize
\begin{tabular}{c|c|c|c|c|c}
Model & $k_p$ & \multicolumn{2}{c|}{Correl. coeff.}  & \multicolumn{2}{c}{RMSE} \\ 
& & $N_t$=100 & $N_t$=469 & $N_t$=100& $N_t$=469\\ 
\hline
BB R0  &  &	0.99963 & 0.96454 & 0.393 & 0.914\\  
BFB tanh & 0 & 0.99966 & 0.96464 & 0.337 & 0.862\\   	 
& 1 & 0.99972 & 0.96496 & 0.288 & 0.818 \\ 	
& 2 & 0.99980 & 0.96556 & 0.193 & 0.748\\
& 2.5 & 0.99984	& 0.96602 & 0.122 & 0.710 \\
& 2.75 & 0.99985 & 0.96629 & 0.084 & 0.696\\
& 3 & 0.99985 & 0.96658	& 0.060 & 0.689\\
& 3.25 & 0.99983 & 0.96688	& 0.082 & 0.693\\
& 3.5 & 0.99979	& 0.96719 & 0.138 & 0.712 \\
& 4	& 0.99960	& 0.96775 & 0.288 & 0.807\\
& 4.5 & 0.99921	& 0.96810 & 0.466 & 0.980\\
& 5	& 0.99856	& 0.96804 & 0.658 & 1.207\\
& 5.5 & 0.99761	& 0.96745 & 0.850 & 1.459\\
& 6	& 0.99641	& 0.96630 & 1.032 & 1.709\\
& 7	& 0.99359	& 0.96271 & 1.335 & 2.141\\
& 8	& 0.99078	& 0.95842 & 1.555 & 2.461\\
& 10	& 0.98642	& 0.95067 & 1.826 & 2.858\\
& 14	& 0.98206	& 0.94157 & 2.073 & 3.222\\
& 30	& 0.97917	& 0.93328 & 2.260 & 3.498\\
& 50	& 0.97784	& 0.92951 & 2.319 & 3.587\\
& 100 & 0.97687	& 0.92600 & 2.408 & 3.716 \\
BB R1.5 & & 0.93959	& 0.86772 & 2.594 & 3.984\\
\end{tabular}
\caption{Root mean square error (RMSE) and correlation coefficient between MD and the BEM-FEM-BEM (BFB) approach using different values of $k_p$ in the tanh function. Out of the 494 tested molecules from the FreeSolv database, we calculated the MSE and correlation coefficient using subsets the $N_t$=100 and $N_t$=469 best ones. BB R0 and R1.5 correspond to coupled BEM-BEM simulations on the SES and a surface 1.5 \AA ~away from the SES, respectively.}
\label{table:ECM}
\end{table}


\subsection*{\sffamily \large Binding energy calculations}



In this section we show results of binding energy calculations of a set of 5 protein-protein and 2 protein-ligand complexes.
As described by Eq. \eqref{eq:bind_energy}, binding energy is the difference between much larger solvation energies, making it a very challenging test in terms of the increased sensibility to meshing and numerical parameters \cite{HarrisBoschitcshFenley2013,harris2015problems,nguyen2017accurate,izadi2018accuracy}.

The protein-protein complexes are 1BRS \cite{buckle1994protein}, 1EMV \cite{kuhlmann2000specificity}, 1JTG \cite{lim2001crystal}, 1BEB \cite{brownlow1997bovine}, and 1A3N \cite{tame2000structures}, and the protein-ligand complexes are 1BBZ \cite{pisabarro1998crystal} and 1SEO \cite{michalska2005crystal}. We downloaded the corresponding PDB and TOP files from the Protein Data Bank\footnote{\url{https://www.rcsb.org/}}, and paramaterized them with the AMBER force field \cite{ponder2003force} using the PDB2PQR software \cite{dolinsky2004pdb2pqr}.
We generated the surface meshes with Nanoshaper \cite{decherchi2013general}, setting the average density to 8 vertices per \AA$^2$, increasing the van der Waals radii by 0, 1.5, and 3 \AA~ in the same way as in Fig \ref{fig:Esfera_CMI} to generate the internal, middle, and external surfaces, respectively. In this case, the TetMesh volumetric mesh had a growth factor of 1.2.


In these simulations the solute's permittivity was \(\epsilon_{m}=2\), while the solvent followed the conditions of water with sodium chloride salt (\(\epsilon_{s}=80\) and \(\kappa=0.125\) ~\AA$^{-1}$). As these structures are larger than the previous results the computational memory and time requirements are much larger. Here, we accelerated the boundary element operators in Bempp-cl using a fast multipole method assembly.


Table \ref{table:1SE0} is another analysis of the importance of the intermediate  surface mesh to generate the volumetric mesh, in this case, for 1SEO and binding energy. As expected, and similar to Fig. \ref{fig:Esfera_CMI}, we see that the effect of the intermediate mesh increases with $k_p$. 
Moreover, since binding free energies are differences of larger solvation energies, the impact is even greater than in the sphere case of Fig. \ref{fig:Esfera_EvsKp}, as we see a divergence of nearly 10\% between using the intermediate mesh and not using it by $k_p$=3, and increasing from there.
All the rest of the results of this section (Table \ref{table:7Proteinas}) use the intermediate mesh.
%
%

\begin{table}[h]
\centering
\footnotesize
\begin{tabular}{c|c|c}
$k_p$ & \parbox{1.6cm}{\centering Without Int. Mesh} & \parbox{1.5cm}{\centering With Int. Mesh}  \\[3mm] 
\hline
0	& -2.227	& -2.293 \\
3 & -13.236	& -12.472 \\
6 & -32.652	& -32.349 \\
9 & -41.453 & -42.759 \\
14 & -45.712	& -49.150 \\
+Inf & -49.934	& -57.458 \\
\end{tabular}
\caption{Binding Energy (in [kcal/mol]) in BEM-FEM-BEM Hyperbolic Tangent for 1SE0 for different intermediate mesh models.}
\label{table:1SE0}
\end{table}


Table \ref{table:7Proteinas} shows binding energy results with \(k_{p}\) values of 6 and 14, which are commonly used in the community, and the $k_p$ that best compared with the experimental energy. 
We acknowledge that this last result may be misleading, as we are not considering the nonpolar component, however, it demonstrates the sensitivity of binding energy with respect to the shape of the variable permittivity on the interface. 
Unlike molecules in the FreeSolv Database, in this case it is more challenging to pinpoint a specific range of \(k_{p}\) that gives the most accurate results.
Even though there are cases where $k_p$ agreed with FreeSolv in Fig. \ref{fig:Mobley} (near $k_p$=3), in others we see that it goes up to values beyond 10.

\begin{table}[h]
\centering
\footnotesize
\begin{tabular}{c|c|c|c}
Structure & $k_p$ & Energy  & Experimental \\ 
\hline
1BBZ & 6 & -2.696 &     \\
     & 14  & -8.746 &   \\
\cline{2-4}
     & 11.50 & -7.977 & -7.99   \\
\hline
1SE0 & 6 & -32.349 &  \\
     & 14 & -49.150 &  \\
\cline{2-4}
     & 2.47 & -9.500 &  -9.50 \\ 
\hline
1BRS & 6 & -53.548 &  \\
     & 14 & -91.313 &  \\
\cline{2-4}
     & 2.17 & 0.919 & 0.96 \\
\hline
1EMV & 6 & -20.376 &   \\
     & 14 & -53.154 &  \\
\cline{2-4}
     & 3.97 & 2.228 & 2.17 \\
\hline  
1JTG & 6 & -9.832 &   \\
     & 14 & -61.147  &  \\
\cline{2-4}
     & 5.37 & 0.417 & 0.40 \\
\hline
1BEB & 6 & 21.274 &   \\
     & 14 & 0.713 &  \\
\cline{2-4}
     & 17.55 & -1.626 &  -1.62 \\
\hline
1A3N & 6 & 61.904 &  \\
     & 14 & -19.493 &  \\
\cline{2-4}
     & 10.57 & 0.155 & 0.16 \\
\end{tabular}
\caption{Binding Energy (in [kcal/mol]) with BEM-FEM-BEM using tanh for different structures.}
\label{table:7Proteinas}
\end{table}

\section*{\sffamily \Large CONCLUSIONS}







In this work, we conducted a systematic analysis of a diffuse interface model of the PB equation for electrostatics in molecular solvation. We focused on a hyperbolic tangent variation of the field parameters, however, the study is applicable to other forms of the diffused interface, like GCS. Our calculations were based on a coupled BEM-FEM-BEM scheme that allowed us to have special treatments of the field parameters in the near-solute region. 

We found that the shape of the hyperbolic tangent in the interfacial region has a large impact on the solvation and binding free energies, and its parameterization (in this case, through the $k_p$ factor) needs to be done carefully. This also means that an appropriate mesh is required, as even though the hyperbolic tangent is continuous, gradients may still be large. 

In solvation energy calculations we saw that a $k_p$ value around 3 yielded the best agreement with MD calculations for the FreeSolv dataset, and ramping $k_p$ up to 10 still gave acceptable accuracy. 
However, binding energies are much more difficult and sensitive. 
In that case, the optimal value ranged from 2 to nearly 20. Such a wide range of values makes it difficult to identify a single best-fit parameter for surface function in binding energy calculations.

We hope this work will inspire new studies of the field parameters near the solute-solvent interface.
On one side, special parameterization for case-dependent $k_p$ could overlook some of the limitations reported here.
On the other side, this BEM-FEM-BEM framework could be used to incorporate field parameters that are extracted from MD calculations or even experiments \cite{fumagalli2018anomalously} accurately, paving the way towards implicit solvent modeling at explicit solvent accuracy.




\subsection*{\sffamily \large ACKNOWLEDGMENTS}
MG-M acknowledges the support from Universidad T\'ecnica Federico Santa Mar\'ia through internal funding for graduate studies. \\
MB acknowledges the support from Kingston University through First Kingston University Grant. \\
CDC acknowledges the support from CCTVal through ANID PIA/APOYO AFB220004.\\

\subsection*{\sffamily \large DATA AVAILABILITY STATEMENT}
All codes to reproduce the results of this manuscript can be found in the GitHub repository \url{https://github.com/bem4solvation/BEM-FEM-BEM_Coupling_with_Tangent_Hiperbolic}.


\clearpage


\bibliography{main}   

\end{document}